# Observation of half-integer level shift of vortex bound states in an iron-based superconductor

Lingyuan Kong[1,2†], Shiyu Zhu[1,2†], Michał Papaj[3†], Lu Cao[1,2], Hiroki Isobe[3], Wenyao Liu[1,2], Dongfei Wang[1,2], Peng Fan[1,2], Hui Chen[1], Yujie Sun[1,4,6], Shixuan Du[1,4], John Schneeloch[5], Ruidan Zhong[5], Genda Gu[5], Liang Fu[3*], Hong-Jun Gao[1,2,4*], and Hong Ding[1,4,6*]

[1]Beijing National Laboratory for Condensed Matter Physics and Institute of Physics, Chinese Academy of Sciences, Beijing 100190, China

[2]School of Physical Sciences, University of Chinese Academy of Sciences, Beijing 100190, China

[3]Department of Physics, Massachusetts Institute of Technology, Cambridge, Massachusetts 02139, USA

[4]CAS Center for Excellence in Topological Quantum Computation, University of Chinese Academy of Sciences, Beijing 100190, China

[5]Condensed Matter Physics and Materials Science Department, Brookhaven National Laboratory, Upton, New York 11973, USA

[6]Songshan Lake Materials Laboratory, Dongguan, Guangdong 523808, China

†These authors contributed equally to this work

*Correspondence to: dingh@iphy.ac.cn, hjgao@iphy.ac.cn, liangfu@mit.edu

**Vortices in topological superconductors host Majorana zero modes (MZMs), which are proposed to be building blocks of fault-tolerant topological quantum computers. Recently, a new single-material platform for realizing MZM has been discovered in iron-based superconductors, without involving hybrid semiconductor-superconductor structures. Here we report on a detailed scanning tunneling spectroscopy study of a $FeTe_{0.55}Se_{0.45}$ single crystal, revealing two distinct classes of vortices present in this system which differ by a half-integer level shift in the energy spectra of the vortex bound states. This level shift is directly tied with the presence or absence of zero-bias peak and also alters the ratios of higher energy levels from integer to half-odd-integer. Our model calculations fully reproduce the spectra of these two types of vortex bound states, suggesting the presence of topological and conventional superconducting regions that coexist within the same crystal. Our findings provide strong evidence for the topological nature of superconductivity in $FeTe_{0.55}Se_{0.45}$ and establish it as an excellent platform for further studies on MZMs.**

Majorana zero modes (MZMs) are proposed to be building blocks of fault-tolerant topological quantum computation[1] due to their non-Abelian statistics. Several systems are predicted to host MZMs, such as intrinsic $p$-wave superconductors[2,3], and multiple heterostructures combining strong spin-orbital coupling (SOC) and superconductivity[4-12]. Recently, a new single-material platform of iron-based superconductors (FeSC) has been discovered[13-15], in which topological nontrivial bands and high-$T_c$ superconductivity coexist naturally[16] without the need of proximity effect common to other proposals. This has led to the observation of a pronounced zero-bias conductance peak (ZBCP) in vortices of FeTe$_{0.55}$Se$_{0.45}$[17] and a related compound[18].

While a ZBCP that does not split across the vortex core is regarded as a strong indication of MZM and topological nature of the superconducting vortex[4,17-19], the observation of ZBCP alone is not enough to prove it. Although several pieces of evidence including spatial profile, tunneling barrier dependence, magnetic field dependence and temperature evolution are fully consistent with MZM in FeTe$_{0.55}$Se$_{0.45}$[17], more convincing verification requires demonstration of the nontrivial topology of the superconducting vortex and underlying band structure. The single crystal of FeTe$_{0.55}$Se$_{0.45}$ is a unique platform to demonstrate the fundamental distinction between the trivial and topological vortices. Its large ratio[17,20] of $\Delta/E_F$ enables realization of the quantum limit[21], where the low-lying quasiparticle bound states, the so-called Caroli-de Gennes-Matricon bound states (CBSs)[22], become discrete levels observable separately within the hard superconducting gap. These bound states are the eigenstates of the vortex planar angular momentum[21-23] with the eigenvalue determined by topological phase of the host superconductor[4,24]. Even though topology dictates existence of two types of discrete bound state spectra, in ordinary circumstances a given material belongs to just one of the classes. This restricts a single sample to either type of the spectrum and thus forbids a direct comparison. However, in a surprising twist, the intrinsic inhomogeneity of FeTe$_{0.55}$Se$_{0.45}$[25], while reducing the number of vortices that host MZM[17], provides a rare opportunity to observe topological and ordinary vortices simultaneously in the same material or even the same region, thus making such a comparison feasible.

Here we report on a systematic scanning tunneling microscopy/spectroscopy (STM/S) study of vortices in FeTe$_{0.55}$Se$_{0.45}$. We observe two topologically distinct classes of vortices, which differ not only by the presence or absence of ZBCP, but also by quantization sequence of the remaining higher energy subgap states. Our detailed analysis compares and contrasts multiple vortices that belong to these two classes and reveals that the ratios of bound state energy value follow either integer or half-odd-integer numbers, provided that the chemical potential is not too small as compared to the superconducting gap. This fundamental difference, arising due to additional angular momentum contribution, is accounted for by our model calculation, which reproduces the discrete bound state spectra and allows us to identify the integer-spaced levels as emerging from topological surface states. In contrast, in an ordinary vortex without MZM, the discrete CBSs energies have half-odd-integer spacing, reflecting the trivial topology of the underlying band structure. This half-integer level shift of vortex bound states between two distinct classes of vortices provides strong evidence for the existence of a pure Majorana zero mode in the FeSC material. Our results also provide a detailed understanding of vortex bound states in FeTe$_{0.55}$Se$_{0.45}$ and in this way facilitate future applications of MZM present in this material platform.

### Integer quantized CBSs in a topological vortex core

To investigate the vortex bound state spectra experimentally, we perform low-temperature ($T_{exp}$ = 0.55 K) high-resolution (0.28 meV) STM/S measurements on as-grown superconducting FeTe$_{0.55}$Se$_{0.45}$ single crystals ($T_c$ = 14.5 K). An atomic resolved lattice structure is observed on the *in-situ* cleaved surface (inset of Fig. 1a). When the magnetic field exceeds $H_{c1}$, superconducting vortices appear as the material enters the mixed state typical of the type-II superconductor[26]. With a 6.0 T magnetic field applied perpendicularly to the sample surface, we find multiple vortices in the zero-bias conductance map shown in Fig. 1a.

In the center of the vortex core, there are sharp ZBCPs with a full width at half-maximum (FWHM) being almost resolution and temperature limited (Extended Data Fig. 1a & 1b). In the previous work[17], we have provided evidence that this ZBCP is a Majorana zero mode induced by the surface Dirac fermions observed in high-resolution angle-resolved photoemission spectroscopy (ARPES) measurements[16]. In this study, we highlight additional high-energy subgap features in the spectrum that are crucial in distinguishing between topological and trivial nature of superconducting vortices and its underlying band structure. To obtain better understanding of the origin of these subgap features, it is beneficial to focus on the vortices where there are several visible peaks inside of the gap. This situation corresponds to vortices present in a region of comparatively smaller, but not too small, $\Delta/E_F$ ratio, to guarantee the presence of several subgap levels with discrete spectra observable under the quantum limit[21]. For such a vortex, the spectrum measured slightly off the center (the blue curve in Fig. 1b) shows three high-energy bound states that coexist with the MZM more clearly. We find that similarly to the MZM, the non-zero energy bound states are also not shifting when changing spatial position of the STM tip (Fig. 1c). The discrete features in the spectrum whose energy does not shift along the real space cut are characteristic of CBSs in the quantum limit[21], as observed previously[27]. The strong electron correlation in this material[28] leads to a large $\Delta/E_F$, thus enabling our experiments well below the required temperature ($T_{exp} < T_{QL} = T_c \Delta / E_F$).

We next examine the level spacing of these discrete CBSs coexisting with a MZM (Fig. 2). We extract the energy positions of each level (Fig. 2c) using a Gaussian fit (Fig. 2b). We identify six discrete levels marked by $L_0$, $L_{\pm1}$, $L_{\pm2}$, $L_{+3}$, with the energy values being 0 meV, 0.65 meV, 1.37 meV, and 1.93 meV, respectively (Extended Data Fig. 1d). It is clear that the CBSs are almost equally spaced in energy. By using the energy of the first level as the energy unit, we present a histogram for each of the levels (Fig. 2d) showing that the ratio of energy levels ($E_L / \Delta E$) closely follows the form of 0 : 1 : 2 : 3. The integer quantized CBSs can be also visualized in an overlap plot (Fig. 2e), with several spatially non-shifting peaks coexisting with the sharp MZM. We summarize the energy level ratios of seven different vortices in which a MZM is observed (Fig. 2f). Although the absolute level energies vary slightly from vortex to vortex (Extended Data Fig. 2), the normalized energy in a unit of first energy level converges to a straight line of integer quantization for all of the vortices present in such regions. It implies that even though in those vortices the CBSs energy values are influenced by local environment, the integer quantized property is robust, as long as the topological nature of underlying band structure remains intact.

We support this conclusion by an energy spectrum calculation using Fu-Kane Model[4,17] (Supplementary Information). We present the comparison between observed peak positions and calculated energy eigenvalues of vortex bound states (Extended Data Fig. 5f). An excellent agreement provides strong evidence for the topological nature of superconductivity in $FeTe_{0.55}Se_{0.45}$, demonstrating that the integer quantized CBS levels are the direct consequence of the topological surface states. In the previous work[17], we focused on vortices with larger level spacing between MZM and first vortex bound state at non-zero energy. Our calculation also reproduces its spectrum precisely when we decrease the value of the chemical potential (Extended Data Fig. 5e). That shows in the case of the chemical potential very close to the Dirac points, a large level spacing will push the first non-zero bound states being very close to the energy of superconducting gap, thus the integer quantization of CBS levels is broken down by quantum confinement effects[29]. We also notice that the energy position of Dirac point is a "sweet spot" for quantum computation. If the chemical potential is exactly located at such a point, a MZM is the only allowed subgap state in a topological vortex core[29-31], and all the other non-zero bound states are pushed to the superconducting gap edge (Supplementary Information).

## Half-odd-integer quantized CBSs in an ordinary vortex core

However, in our samples there exists another class of vortices that do not contain MZMs. To examine their origin, we perform a comparison study for the CBSs in these ordinary vortices. Similarly as in the case of a topological vortex, the CBSs in the ordinary vortex are discrete in energy (Fig. 3). The first CBS level ($L_{+1}$) is located at 0.26 meV. The energies of higher levels ($L_{+2}$, $L_{+3}$, $L_{+4}$, $L_{+5}$) are found to be 0.83 meV, 1.34 meV, 1.84 meV, 2.34 meV, respectively. These CBSs show a strong particle-hole asymmetry, being strong in the positive energy and very weak in the negative energy (Fig. 3b). The particle-hole asymmetry is a common phenomenon for the superconducting vortex core for FeSC materials[32,33], though the degree of asymmetry varies for different vortices (Extended data Fig. 4).

Although we cannot locate the $L_{-1}$ level of CBSs in Fig. 3, the absence of the ZBCP indicates that no bound state in the vortex has angular momentum eigenvalue equal to zero. By using the first level spacing as a unit, we summarize the ratios ($E_L / \Delta E$) of the three vortices in Fig. 3d. Despite a strong variation of particle-hole asymmetry among these vortices, the ratios converge into a straight line of half-odd-integer quantization with the form of 0.5 : 1.5 : 2.5 : 3.5 : 4.5. The appearance of CBSs with energy values proportional to half-odd-integers in a vortex without a ZBCP is consistent with the expected behavior of an ordinary vortex core in which only the pairing in conventional bands contributes to quasiparticle excitations under a magnetic field[21-23,34]. The angular momentum eigenvalues of bound states in an ordinary vortex are half of an odd integer. Accordingly, the energies of CBSs inherit the half-odd-integer quantization with an equal level spacing close to the center of the gap. In this case we also provide numerical calculation of the energy spectrum based on solving Bogoliubov-de Gennes equation with the parabolic conventional bands that reproduces the experimental energy values (Extended Data Fig. 3).

## Characteristic spatial pattern of the quantized CBSs

Friedel-like oscillation of local density of states has been predicted in half-odd-integer quantized CBSs of an ordinary vortex core, with the spatial periodicity being approximately of the scale of Fermi wave length $\sim 1/k_F$ [21,35]. The typical $k_F$ of conventional bands is larger than 0.1 Å$^{-1}$ [36], leading to the spatial oscillation of CBSs within 1 nm in an ordinary vortex core, which is difficult to be observed by STM[26,27, 32-34]. However, in FeTe$_{0.55}$Se$_{0.45}$, a minimal value of $k_F$ of the order of 0.01 Å$^{-1}$ was observed for its Dirac surface state[16]. Therefore, the resulting large oscillation periodicity enables easier observation of the spatial pattern of CBSs by STM. As a final piece of evidence for topological nature of the vortices that contain ZBCP, we perform a constant-bias conductance mapping of the three lowest levels of the integer-quantized CBSs (Fig. 4). While the ZBCP and the first-level CBS ($L_{+1}$) display a solid circle spatial pattern around the center of vortex core (left panels of Figs. 4a, b), the second-level CBS ($L_{+2}$) shows a hollow ring pattern around the vortex center (left panel of Fig. 4c). This pattern is unique to Dirac fermions of topological surface states with spin-momentum locking, whereas in ordinary vortices only a single bound state has a wave function maximum at the center of the vortex[21,38]. The measurement is also fully consistent with our numerical calculation (middle panels of Figs. 4a-c) which clearly reveals this spatial pattern difference of the wave functions of these three levels (right panels of Figs. 4a-c).

## Half-integer level shift between two classes of vortices

We have clearly observed the distinction in the energy spectra of vortex bound states in topological and ordinary vortices. In an ordinary vortex core (Fig. 5a), only the conventional bands contribute to quasiparticle excitations and the bound states have eigenvalues of angular momentum that are half-odd-integer as a result of addition of integer orbital contribution $L$ and half-odd-integer vorticity contribution $\tau$ (for vortices with an odd winding number). Accordingly, the energy eigenvalues of CBSs are also approximately half-odd-integer quantized, i.e. $E_\nu = \nu \Delta^2/E_F$ ($\nu = \pm 1/2, \pm 3/2, \pm 5/2, \ldots$), with $\nu$ being the eigenvalue of angular momentum[21-23]. On the other hand, topological vortices (Fig. 5b) that arise due to superconductivity in Dirac surface states gain additional half-odd-integer contribution $S$ to angular momentum due to intrinsic spin carried by Dirac fermions[4,29,37]. This leads to a half-integer shift of the angular momentum, thus being an integer and the energy values of the bound states become integer quantized, i.e. $E_\nu = \nu \Delta^2/E_F$ ($\nu = 0, \pm 1, \pm 2, \pm 3, \ldots$). Majorana zero modes can then be regarded as a special zero-energy CBS for a topological superconducting state with $\nu = 0$, as long as the zero energy CBS is equal-weight mixture of particle/hole components, and the spin degree of freedom is frozen out[24].

## Summary and outlook

By means of direct comparison between topological and ordinary vortices in the same crystal of $FeTe_{0.55}Se_{0.45}$, we clearly demonstrated half-integer level shift of CBSs around a Majorana zero mode. This indicates that these two types of vortices are in different topological phases[24], varying with the participation of Dirac surface states in superconducting quasiparticle excitations[4,29,37]. The larger level spacing between the MZM and first CBS in this material, especially when the material approaches the zero doping limit[29-31] (Extended Data Fig. 6), protects the MZM from external perturbations[38], which is favorable for demonstration of non-Abelian statistics of MZMs in a braiding operation[1].

1. Nayak, C. *et al*. Non-Abelian anyons and topological quantum computation. *Rev. Mod. Phys.* **80**, 1083–1159 (2008).
2. Read, N. *et al*. Paired states of fermions in two dimensions with breaking of parity and time-reversal symmetries and the fractional quantum Hall effect. *Phys. Rev. B* **61**, 10267–10297 (2000).
3. Kitaev, A. Y. Unpaired Majorana fermions in quantum wires. *Phys. Uspekhi* **44,** 131–136 (2001).
4. Fu, L. & Kane, C. L. Superconducting proximity effect and Majorana fermions at the surface of a topological insulator. *Phys. Rev. Lett.* **100,** 096407 (2008).
5. Lutchyn, R. M., Sau, J. D. & Das Sarma, S. Majorana fermions and a topological phase transition in semiconductor-superconductor heterostructures. *Phys. Rev. Lett.* **105,** 077001 (2010).
6. Oreg, Y., Refael, G. & von Oppen, F. Helical liquids and Majorana bound states in quantum wires. *Phys. Rev. Lett.* **105,** 177002 (2010).
7. Mourik, V. *et al*. Signatures of Majorana fermions in hybrid superconductor-semiconductor nanowire devices. *Science* **336,** 1003–1007 (2012).
8. Nadj-Perge, S. *et al*. Observation of Majorana fermions in ferromagnetic atomic chains on a superconductor. *Science* **346,** 602–607 (2014).
9. Sun, H. H. *et al*. Majorana zero mode detected with spin selective Andreev reflection in the vortex of a topological superconductor. *Phys. Rev. Lett.* **116,** 257003 (2016).
10. Deng, M. T. *et al*. Majorana bound state in a coupled quantum-dot hybrid-nanowire system. *Science* **354,** 1557–1562 (2016).
11. Jeon, S. *et al*. Distinguishing a Majorana zero mode using spin-resolved measurements. *Science* **358**, 772–776 (2017).
12. Zhang, H. *et al*. Quantized Majorana conductance. *Nature* **556**, 74–79 (2018).
13. Wang, Z.-J. *et al*. Topological nature of the $FeSe_{0.5}Te_{0.5}$ superconductor. *Phys. Rev. B* **92**, 115119 (2015).
14. Wu, X.-X. *et al*. Topological characters in $Fe(Te_{1-x}Se_x)$ thin films. *Phys. Rev. B* **93**, 115129 (2016).
15. Xu, G. *et al*. Topological superconductivity on the surface of Fe-based superconductors. *Phys. Rev. Lett.* **117**, 047001 (2016).
16. Zhang, P. *et al*. Observation of topological superconductivity on the surface of iron-based superconductor. *Science* **360**, 182–186 (2018).
17. Wang, D. *et al*. Evidence for Majorana bound states in an iron-based superconductor. *Science* **362**, 333 (2018).

18. Liu, Q. et al. Robust and clean Majorana zero mode in the vortex core of high-temperature superconductor $(Li_{0.84}Fe_{0.16})OHFeSe$. https://arxiv.org/abs/1807.01278 (2018).
19. Xu, J.-P. et al. Experimental detection of a Majorana mode in the core of a magnetic vortex inside a topological insulator-superconductor $Bi_2Te_3/NbSe_2$ heterostructure. *Phys. Rev. Lett.* **114**, 017001 (2015).
20. Rinott, S. et al. Tuning across the BCS-BEC crossover in the multiband superconductor $Fe_{1+y}Se_xTe_{1-x}$: An angle-resolved photoemission study. *Sci. Adv.* **3**, e1602372 (2017).
21. Hayashi, N. et al. Low-Lying Quasiparticle excitations around a vortex core in quantum limit. *Phys. Rev. Lett.* **80**, 2921 (1998).
22. Caroli, C., de Gennes, P. G. & Matricon, J. Bound fermion states on a vortex line in a type II superconductor. *Phys. Lett.* **9**, 307–309 (1964).
23. Gygi, F. & Schluter, M. Self-consistent electronic structure of a vortex line in a type-II superconductor. *Phys. Rev. B* **43**, 7609 (1990).
24. Alicea, J. New directions in the pursuit of Majorana fermions in solid state systems. *Rep. Prog. Phys.* **75** 076501 (2012).
25. Massee, F. et al. Imaging atomic-scale effects of high-energy ion irradiation on superconductivity and vortex pinning in Fe(Se,Te). *Sci. Adv.* **1**, e1500033 (2015).
26. Suderow, H. et al. Imaging superconducting vortex cores and lattices with a scanning tunneling microscope. *Supercond. Sci. Technol.* **27**, 063001 (2014).
27. Chen, M. et al. Discrete energy levels of Caroli-de Gennes-Martricon states in quantum limit in $FeTe_{0.55}Se_{0.45}$. *Nat. Commun.* **9**, 970 (2018).
28. Yin, Z. P., Haule, K. & Kotliar, G. Spin dynamics and orbital-antiphase pairing symmetry in iron-based superconductors. *Nat. Phys.* **10**, 845–850 (2014).
29. Khaymovich, I. M. et al. Vortex core states in superconducting graphene. *Phys. Rev. B* **79**, 224506 (2009).
30. Jackiw, R. & Rossi, P. Zero modes of the vortex-fermion system. *Nucl. Phys. B* **190**, 681 (1981).
31. Ghaemi, P. & Wilczek, F. Near-zero modes in superconducting graphene. *Phys. Scr.* **146**, 014019 (2012).
32. Shan, L. et al. Observation of ordered vortices with Andreev bound states in $Ba_{0.6}K_{0.4}Fe_2As_2$. *Nat. Phys.* **7**, 325–331 (2011).
33. Hanaguri, T. et al. Scanning tunneling microscopy/spectroscopy of vortices in LiFeAs. *Phys. Rev. B* **85**, 214505 (2012).
34. Hess, H. F., Robinson, R. B. & Waszczak, J. V. Vortex-core structure observed with a scanning tunneling microscopy. *Phys. Rev. Lett.* **64**, 2711–2714 (1990).
35. Kaneko, S.-I. et al. Quantum limiting behaviors of a vortex core in an anisotropic gap superconductor. *J. Phys. Soc. Jpn.* **81**, 063701 (2012).
36. Coldea, A. I. & Watson, M. D. The key ingredients of the electronic structure of FeSe. *Annu. Rev. Condens. Matter Phys.* **9**, 125–146 (2018).
37. Hu, L.-H. et al. Theory of spin-selective Andreev reflection in the vortex core of a topological superconductor, *Phys. Rev. B* **94**, 224501 (2016).
38. Colbert, J. & Lee, P. A. Proposal to measure the quasiparticle poisoning time of Majorana bound states. *Phys. Rev. B* **89**, 140505 (2014).

**Acknowledgements** We thank Noah. F. Yuan, Ching-Kai Chiu, Constantin Schrade for helpful discussions, and Fazhi Yang, Guojian Qian for technical assistance. This work at IOP is supported by grants from the Ministry of Science and Technology of China (2015CB921000, 2015CB921300, 2016YFA0202300), the National Natural Science Foundation of China (11234014, 11574371, 61390501), and the Chinese Academy of Sciences (XDB28000000, XDB07000000). L.F. and G.D.G are supported by US DOE (DE-SC0010526, DE-SC0012704, respectively). J.S. and R.D.Z. are supported by the Center for Emergent Superconductivity, an EFRC funded by the US DOE.

**Author Contributions** H.D. and H.-J.G. designed the experiments. S.-Y.Z. and L.C. performed the STM experiments with assistance of L.-Y.K., W.-Y.L., D.-F.W., P.F., H.C. and S.-X.D. M.P., H.I. and L.F. provided theoretical models and simulations. J.S., R.-D.Z. and G.-D.G. provided samples. L.-Y.K., S.-Y.Z. and H.D. analyzed experiment data with inputs from all other authors. L.-Y.K., M.P. and S.-Y. Z plotted figures with inputs from all other authors. All the authors participated in writing the manuscript. H.D., H.-J.G. and L.F. supervised the project.

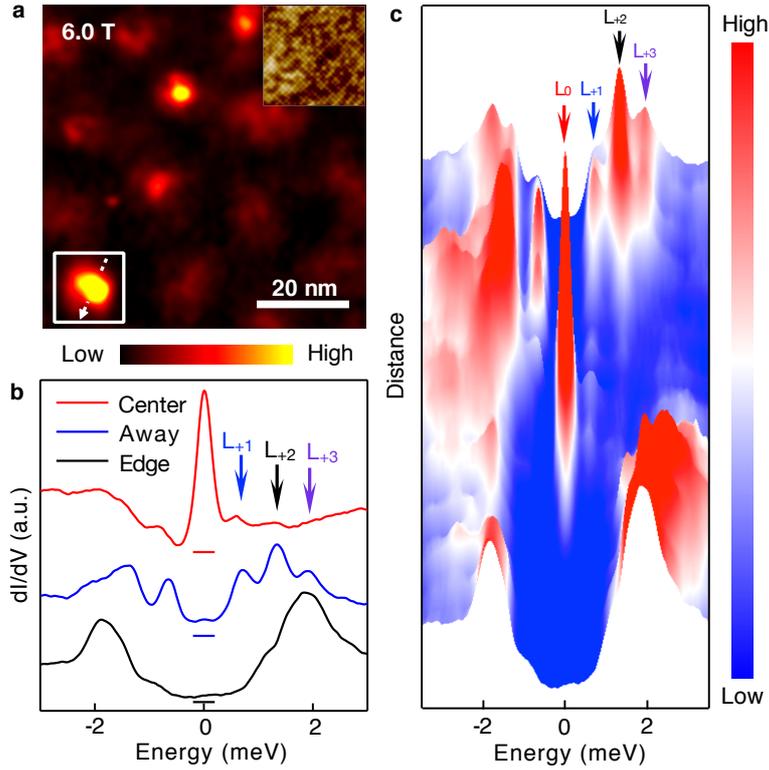

**Figure 1 | Caroli-de Gennes-Matricon states in a vortex with a Majorana zero mode. a,** A normalized zero-bias conductance map measured at a magnetic field of 6.0 T, with the area 70 nm by 70 nm. The average inter-vortex length is around 19.3 nm, which is consistent with the expectation, $L = \sqrt{2\Phi_0/\sqrt{3}B}$. Insert: A STM topography of FeTe$_{0.55}$Se$_{0.45}$ (scanning area 10 nm by 10 nm). **b,** Typical tunneling conductance spectra measured around the vortex marked by the white box in **(a)**. The curves are offset for clarity. The red curve is measured at the vortex center. The blue curve is measured slightly off the center, and the black curve is measured at the vortex edge. The short colored bar below each curve marks its zero conductance. **c,** Three dimensional display of the line-cut intensity plot along the white dash line indicated in **(a)**. Four subgap states are identified by the arrows in different colors. Besides the MZM, all the Caroli-de Gennes-Matricon states almost remain at the same energy along the line cut through the vortex.

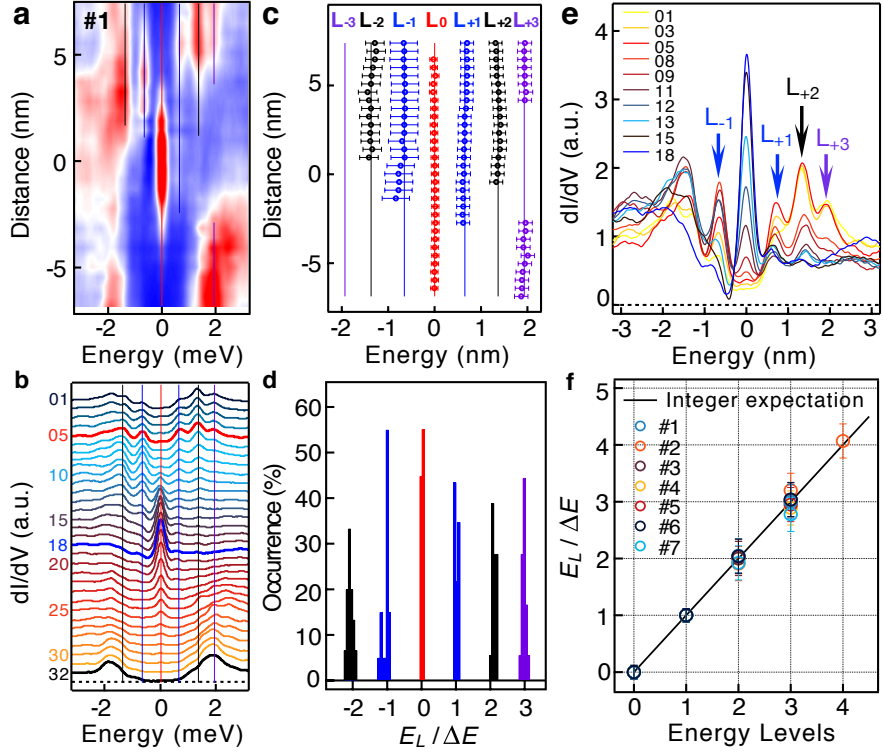

**Figure 2 | Integer quantized CBSs in a topological vortex. a,** A line-cut intensity plot of the topological vortex #1, discrete quantized CBSs are marked by colored solid lines. **b,** A waterfall-like plot of **(a)** with 32 spectra. Spectra numbers are marked on the left side. The curves in red, blue and black are shown in **Fig. 1(b)** with same colors. **c,** The energy values of the observed CBSs at different spatial positions. The energies of CBSs, marked by $L_0$, $L_{\pm 1}$, $L_{\pm 2}$, $L_{\pm 3}$, are extracted from **(b)** with a Gaussian fit. The error bar is calculated by the standard deviation of each energy level. The standard deviation of the MZM is 0.08 meV, smaller than the lock-in modulation energy used in the experiments, $eV_{mod} = 0.1$ meV. The energy values of the solid lines are calculated by the average energy of each level. **d,** A histogram of the energy values of all the observed subgap states. The sampling width is 40 μeV. The energy of the horizontal axis is normalized by the first level spacing, i.e. the ratio $E_L/\Delta E$. **e,** An overlapping plot of 10 $dI/dV$ spectra selected in **(b)**. **f,** Summary of $E_L/\Delta E$ vs level number data for topological vortices #1 to #7. It demonstrates that the CBSs energies are proportional to integers. The solid line is calculated using $E_L/\Delta E = E_L/(\Delta^2/E_F) = \nu$, $\nu$ being the number of energy level.

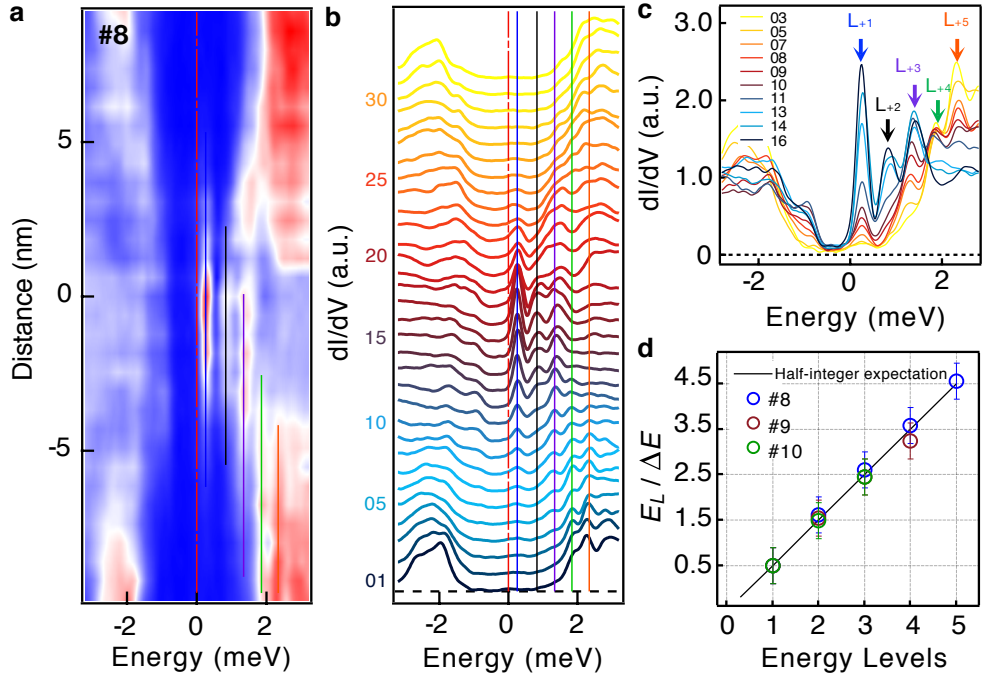

**Figure 3 | Half-odd-integer quantized CBSs in an ordinary vortex. a,** A line-cut intensity plot of the ordinary vortex #8, the CBSs are marked by solid lines with colors. **b,** A waterfall-like plot of (a) with 32 spectra. Spectra numbers are marked on the left side. **c,** An overlapping plot of 10 $dI/dV$ spectra selected in (b) with each energy level of the CBSs marked by $L_{+1}$, $L_{+2}$, $L_{+3}$, $L_{+4}$, $L_{+5}$ on the top. **d,** Summary of $E_L/\Delta E$ vs level number data for ordinary vortices #8 to #10. $\Delta E$ is determined by double of energy of the lowest level. It demonstrates that the CBSs energies are proportional to half-odd-integers. The solid line is calculated using $E_L/\Delta E = E_L/(\Delta^2/E_F) = (2\nu-1)/2$, $\nu$ being the number of energy level.

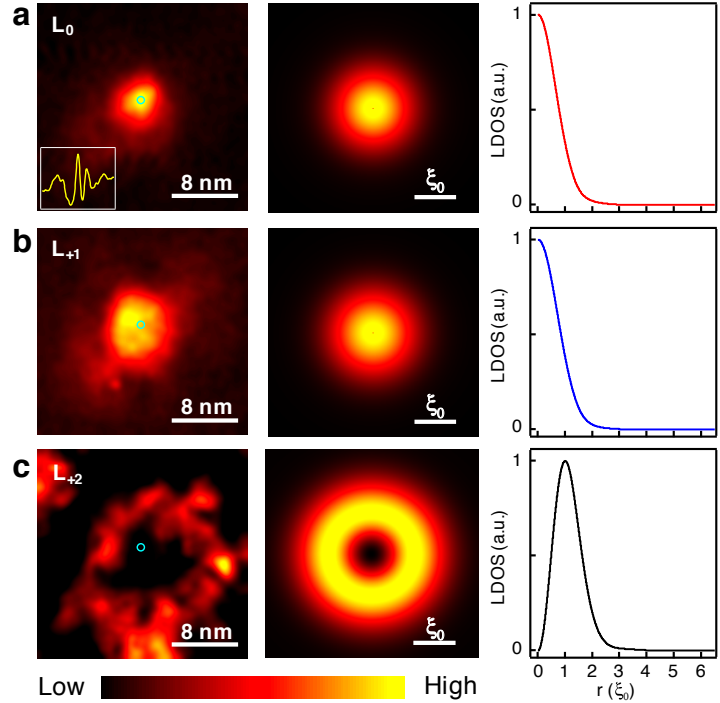

**Figure 4 | Spatial pattern of integer quantized CBSs. a - c,** Comparison plots between STM measurements and numerical calculations for $L_0$ (MZM), $L_{+1}$, $L_{+2}$, respectively. The insert curve in **(a)** is a typical STM spectrum measured at a vortex center, that integer-quantized CBS levels are located at 0.6 meV, 1.3 meV, and 1.8 meV for $L_{+1}$, $L_{+2}$ and $L_{+3}$, respectively. The first column is the constant bias conductance maps of each CBS levels with the area 25 nm by 25 nm, the cyan symbols marked on the three images are the locations of the vortex centers extracted from **(a)**. The MZMs ($L_0$) and the first CBS level ($L_{+1}$) have the strongest intensity at the vortex center, however, the second CBS level shows a ring-like feature around the vortex center with an offset. The radius of the ring ($R$) is about 7 nm, which corresponds to the value of $k_F \sim 1/R \sim 0.014$ $A^{-1}$, similar to the results of ARPES measurements on the Dirac surface states. The second column shows numerical calculations of two-dimensional local density of states based on the model we constructed to simulate a topological vortex core (Supplementary Information). The third column displays the radial wave function of each level.

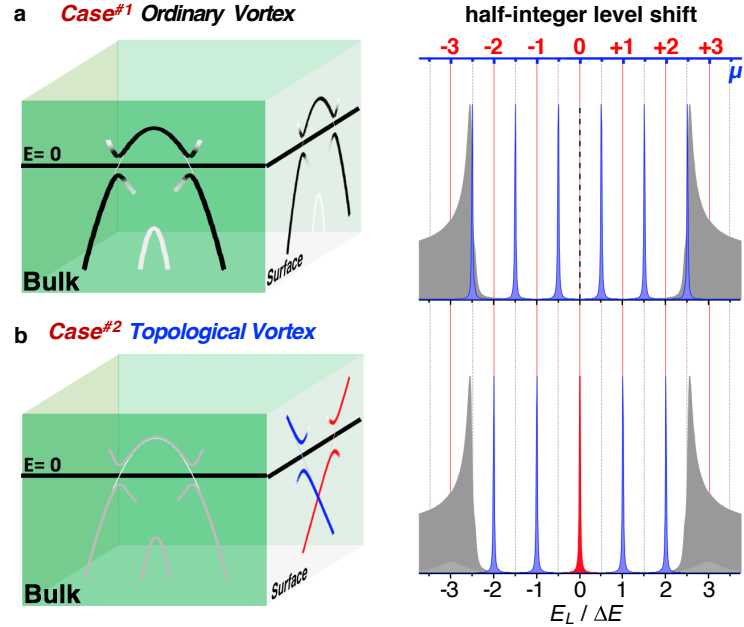

**Figure 5 | Half-integer level shift around a Majorana zero mode.** First column: Schematic plots of the underlying surface and bulk band structure within different cases. Second column: Schematic plots of the subgap CBSs. The blue axis marks the eigenvalues of vortex planar angular momentum. **a,** When the underlying band structure is topological trivial, the vortices behave ordinarily. The bulk bands with spin degeneracy dominate quasiparticle excitations inside the vortex. Due to the absence of the Dirac electrons on sample surface, quasiparticles can only feel the phase winding of the ordinary superconducting vortex, which leads to half-odd-integer angular momenta, and related half-odd-integer quantized CBSs. **b,** When the underlying band structure is dominated by the topological surface state, vortex quasiparticle excitations gain an additional angular momentum from the Dirac electrons. It induces a half-integer level shift of those CBSs as compared with **(a)**, the zero-energy bound state becomes a Majorana zero mode, due to the effective spinless p-wave-like pairing induced on the Dirac surface states. If the chemical potential in **(b)** is tuned to the Dirac point, which is the zero doping limit, all the CBSs are pushed towards the gap edge, leaving the MZM isolated at zero energy (Extended Data Fig. 6).

# METHODS

**Sample preparation.** High quality single crystals of $FeTe_{0.55}Se_{0.45}$ were grown using the self-flux method, and their values of $T_c$ were determined to be 14.5 K[39] from magnetization measurements. There are two kinds of single crystals crystallizing simultaneously with similar structures and Te/Se compositions. $Fe_{1+y}Te_{0.55}Se_{0.45}$ single crystals with excess Fe atoms, with shinning surfaces that are easy to cleave, are non-superconducting before annealing under Te atmosphere. $FeTe_{0.55}Se_{0.45}$ single crystals without excess Fe, usually without shinning surface, are superconducting without post-annealing. All STM/STS data shown in this paper are from as-grown $FeTe_{0.55}Se_{0.45}$ single crystals[17].

**STM measurements.** The samples used in the experiments were cleaved *in-situ* and immediately transferred to a STM head. Experiments were performed in two different ultrahigh vacuum ($1 \times 10^{-11}$ mbar) LT-STM systems, STM#1 (USM-1300s-$^3$He) and STM#2 (USM-1300-$^3$He with a vector magnet)[17], STM images were acquired in the constant-current mode with a tungsten tip. Differential conductance (*dI/dV*) spectra were acquired by a standard lock-in amplifier at a frequency of 973 Hz, under modulation voltage $V_{mod}$ = 0.1 mV, set point voltage $V_s$ = -5 meV and tunneling current $I_t$ = 200 pA. The equipment energy resolution was calibrated on a clean Nb (110) surface, being 0.27 meV for STM#1 and 0.23 meV for STM#2. The voltage offset calibration was followed by a standard method of overlapping points of I-V curves. Low temperatures of 0.4 K can be achieved by a single-shot $^3$He cryostat. A perpendicular magnetic field up to 11 Tesla for STM#1 and a vector magnetic field with the maximum value $9_z$-$2_x$-$2_y$ Tesla for STM#2 can be applied to a sample surface. All the data were measured by STM#2, except the data shown in Extended Data Fig. 2d.

**Tunneling barrier dependence of ZBCP within integer quantized CBSs.** To check the Majorana behavior of ZBCPs within integer quantized CBSs, we performed tunneling barrier evolution measurements on vortex #1. The ZBCPs do not shift across two orders of magnitude of tunneling barrier conductance (Extended Data Fig. 1a). FWHMs of ZBCPs, extracted by a simple Gaussian fit, are about 0.316 meV, which also barely change across two orders of magnitude of tunneling barrier conductance. We calculate the background of vortex #1, that defined as integrated area from -1 meV to +1 meV of the spectrum measured at vortex edge[17]. The background of vortex #1 is about 0.51, which is consistent with the scenario proposed previously[17], that large background conductance induces extra energetic broadening of ZBCP.

**Possible origin of the two classes of superconducting vortices.** The spatial profile of superconducting order parameter is anti-correlated with the local density of states, which enables the detection of the vortices in a STM/S measurement[26]. We found that topological and ordinary vortices can coexist in the same sample, and even can coexist in the same area within several hundred nanometers (Fig. 1a). One of the possible explanations is that the competition between Dirac surface states and conventional bulk band in quasiparticle excitations determines topological phase of vortices. Considering the intrinsic chemical inhomogeneity in such Telluride/Selenide alloy, it is reasonable that the competition is different among spatial positions. To be explicit, high resolution ARPES measurements[16] shows the bulk SOC gap is about 20 meV, which is one order of magnitude smaller than that in $Bi_2Se_3$ family. Generally, Dirac surface states are protected by the time-reversal symmetry, so that they cannot be destroyed by weak perturbations from non-magnetic scattering[40]. Even though such a mechanism truly holds in $FeTe_{0.55}Se_{0.45}$, there is another possibility originating from the small SOC gap to break down topological protection. Any kinds of scattering can destroy topological nontrivial properties by overcoming the small SOC gap, which protects Dirac surface state. Those scattering potentials could be introduced by chemical disorder or multiple kinds of impurities beneath the sample surface. As long as the potential strength is larger than 20 meV, in the vicinity of the scattering potential, Dirac surface states are absent[41,42], thus vortex excitations are dominated by conventional bands, and half-odd-integer quantized CBSs show up. That provides a clue that observation of MZM in $FeTe_{0.55}Se_{0.45}$ requires ultra-high sample qualities with low scattering strength. This idea is also supported by the recent experiments. The study of $FeTe_{0.55}Se_{0.45}$[27], which only found half-odd-integer quantized CBSs, shows a disordered vortex configuration, while in our previous study which observed sharp MZMs[17], shows ordered vortex lattice (also see Extended Data Fig. 1c). Moreover, vortex bound states are also studied on a stoichiometric FeSe-plane of $Li_{0.84}Fe_{0.16}OHFeSe$[18]. It was claimed that MZMs can be found in all impurity-free vortices, while in defect pinned vortices ZBCP disappears. The break-down mechanism of topological band structure proposed here is a combination of disordered scattering potential and small topological bulk gap, which potentially leads to MZMs being found in a fraction of vortices. Even though the presence of topological surface states is spatially nonuniform, the topological origin of surface Dirac fermions is still universal and robust[43].

**Toward the zero doping limit of a topological vortex core.** Occasionally, in some vortices we observe a strong MZM peak at the zero energy without other subgap states in its vicinity, with one example shown in Extended Data Fig. 5, which is the same as the one in our previous study[17]. By carefully examining the spectra, we find three pairs of subgap states at the higher energies near the gap edge. To visualize the subgap state more clearly, we display several spectra measured at around -5 nm of Extended Data Fig. 5a, together with a gray curve measured far away from the vortex center showing the full superconducting gap (Extended Data Fig. 5b). It shows clearly that on the negative energy, there are three energy levels marked by $L_{-1}$, $L_{-2}$, $L_{-3}$, in which the peaks are barely shifted at different spatial positions. This indicates those subgap features are also quantized CBSs. The average energy of each level is 1.12 meV, 1.65 meV and 2.06 meV, respectively. It neither satisfies integer-quantization nor half-odd-integer quantization. In order to get a detailed understanding of the difference between the vortex #1 shown in Figs. 1&2 and the vortex #11 shown in Extended Data Fig. 5, we display their normalized intensity line profile at zero energy in Extended Data Fig. 5c. It is evident that the line profile of vortex #11 has a broader distribution, while the other two vortices are more concentrated around the vortex center. We implement Fu-Kane Model[4,17] to simulate this difference qualitatively. As shown in Extended Data Fig. 5d, it shows the smaller Fermi energy of the Dirac band corresponding to the wider spatial distribution of the MZM wave function.

The vortex quasiparticle spectrum of Dirac fermions has been studied in particle physics previously[30,31]. The subgap spectra of a vortex under zero-doping limit, that is chemical potential exactly located at charge neutrality (Dirac) point, have *n* zero energy states isolated (n being the winding number of superconducting vortex). All the other higher energy states are pushed to the superconducting gap edge $\pm|\Delta_0|$. That provides an optimal condition for observing and further manipulating MZMs (Extended Data Fig. 6). If the chemical potential slightly deviates from zero doping limit, but with a sufficient small $E_F$, the integer quantization of CBS is still absent due to quantum confinement effects[29]. Non-zero CBSs will crowd at higher energies close to the superconducting gap edge, and leave the MZM nearly isolated at the zero energy. In order to confirm this expectation quantitatively, we performed a numerical simulation on a Majorana vortex core. By using a very small Fermi energy, we reproduced the results of vortex #11 (Extended Data Fig. 5e).

**Numerical calculation.** To model the experimental results, we performed calculations using Bogoliubov-de Gennes (BdG) Hamiltonian for 2D surface states with both linear (Dirac) and parabolic dispersion. We then assumed that a vortex with vorticity $|n| = 1$ is placed at the origin, which corresponds to the spatial dependence of the superconducting order parameter given by $\Delta(r,\phi) = \Delta_0 f(r) e^{i\phi}$. The specific form of the radial dependence $f(r)$ is fitted for each case separately. For the Dirac surface states, the BdG Hamiltonian is:
$$H_{Dirac} = v_F \tau_z (p_x \sigma_x + p_y \sigma_y) - \mu \tau_z + \Delta(r,\phi) \tau_x$$
$\tau$ and $\sigma$ are Pauli matrices describing the particle-hole and spin spaces, respectively, $v_F$ is the Fermi velocity and $\mu$ is the chemical potential. Since we have assumed rotational symmetry of the vortex, we can express the BdG equations as a set of 1D radial equations separated into angular momentum modes. We therefore use the following ansatz:
$$\psi(r,\phi) = \frac{e^{iv\phi - \frac{i\phi}{2}\sigma_z + i\frac{\pi}{4}\sigma_z + \frac{i\phi}{2}\tau_z}}{\sqrt{r}} \begin{pmatrix} u_\uparrow(r) \\ u_\downarrow(r) \\ v_\downarrow(r) \\ v_\uparrow(r) \end{pmatrix}$$

where $\nu$ is the angular momentum. We can express all the lengths in terms of the coherence length $\xi = \frac{\hbar v_F}{\Delta_0}$ and all the energies in terms of $\Delta_0$ to finally obtain the set of differential equations:

$$\begin{pmatrix} -\bar{\mu} & -\frac{d}{d\bar{r}} - \frac{\nu + \frac{1}{2}}{\bar{r}} & f(\bar{r}) & 0 \\ \frac{d}{d\bar{r}} - \frac{\nu + \frac{1}{2}}{\bar{r}} & -\bar{\mu} & 0 & f(\bar{r}) \\ f(\bar{r}) & 0 & \bar{\mu} & \frac{d}{d\bar{r}} + \frac{\nu - \frac{1}{2}}{\bar{r}} \\ 0 & f(\bar{r}) & -\frac{d}{d\bar{r}} + \frac{\nu - \frac{1}{2}}{\bar{r}} & \bar{\mu} \end{pmatrix} \begin{pmatrix} u_\uparrow(\bar{r}) \\ u_\downarrow(\bar{r}) \\ v_\downarrow(\bar{r}) \\ v_\uparrow(\bar{r}) \end{pmatrix} = E \begin{pmatrix} u_\uparrow(\bar{r}) \\ u_\downarrow(\bar{r}) \\ v_\downarrow(\bar{r}) \\ v_\uparrow(\bar{r}) \end{pmatrix}$$

with $\bar{\mu} = \mu/\Delta_0$ and $\bar{r} = \frac{r}{\xi}$. This set of equations is then discretized on a 1D lattice (equivalent to solving the equations on a disk with radius $R = 100\,\xi$) and lowest lying eigenvalues and eigenvectors are obtained. To avoid fermion doubling problem, we use the approach of Susskind adjusted to quasi-1D radial geometry[44-46]. The eigenvectors are then used to calculate the local density of states (presented in Fig. 4 of the main text) by using:

$$LDOS(E, r) = \frac{1}{r} \sum_{n, \sigma = \uparrow, \downarrow} |u_{n,\sigma}(r)|^2 \delta(E - E_n) + |v_{n,\sigma}(r)|^2 \delta(E + E_n)$$

where the sums are taken over the positive eigenvalues and the spin components. The 2D density maps are then obtained from the radial dependence by using the rotational symmetry of the wave functions in our model.

We also modeled the vortices that appear in the usual parabolic bands, which do not host Majorana fermions. The BdG Hamiltonian is then:

$$H_{par} = \left(\frac{p^2}{2m} - \mu\right) \tau_z + \Delta(r, \phi) \tau_x$$

where again $\tau$ are Pauli matrices in particle-hole space and $m$ is the effective mass of the band. We similarly decompose the BdG equations into angular momentum modes, obtaining in this a way a set of 1D radial equations. With energies expressed in units of $\Delta_0$ and lengths in units of $\xi$, the equations read:

$$\begin{pmatrix} -\frac{1}{4\bar{\mu}}\left(\frac{d^2}{d\bar{r}^2} + \frac{1}{\bar{r}}\frac{d}{d\bar{r}} - \frac{\left(\nu + \frac{1}{2}\right)^2}{\bar{r}^2}\right) - \bar{\mu} & f(\bar{r}) \\ f(\bar{r}) & \bar{\mu} + \frac{1}{4\bar{\mu}}\left(\frac{d^2}{d\bar{r}^2} + \frac{1}{\bar{r}}\frac{d}{d\bar{r}} - \frac{\left(\nu - \frac{1}{2}\right)^2}{\bar{r}^2}\right) \end{pmatrix} \begin{pmatrix} u(\bar{r}) \\ v(\bar{r}) \end{pmatrix} = E \begin{pmatrix} u(\bar{r}) \\ v(\bar{r}) \end{pmatrix}$$

Then the equations are discretized on a 1D lattice. To avoid large numerical errors for small angular momentum values, we employed the discretization scheme proposed in Ref. 47.

In order to obtain better fitting of the eigenvalues to the experimental results, we modified the radial profile of the vortex $f(r)$ from the standard form of $\tanh\frac{r}{\xi}$. Since the iron-based superconductors have a rich phenomenology, there may be multiple causes responsible for such a change. Among these are the effects of multiple bands and gaps or in the case of the alloys (like the material under consideration) inhomogeneities of composition that can lead to spatially dependent superconducting order parameter even in zero field. To model such effects, we introduce a phenomenological radial profile of the vortex, which we obtain by multiplying the standard form by the spatial profile of the presumed region of suppressed superconductivity:

$$f(r) = \frac{\left(1 + \Delta_{min} + (1 - \Delta_{min})\tanh\frac{r - b\,\xi}{c\,\xi}\right)}{2}\tanh\frac{r}{\xi}$$

We then keep $\Delta_{min} = \frac{1}{2}$ and for each case adjust the values of $b$ and $c$. The radial profiles for the three cases highlighted in the paper are presented in Extended Data Fig. 7a. In Extended Data Fig. 7b, we present the comparison of eigenvalue fitting with the standard and the modified vortex core profiles. In the standard form of the radial function, the only parameter is the chemical potential and the fit uses the least squares method. While the results with the standard vortex profile are mostly within the error bars of the experimental data, the ratios of energies differ from the measured values. However, once we perform the calculation using the modified profile, the agreement with the experimental values improves significantly and the ratios are now correctly reproduced.

39. Wen, J. *et al.* Short-range incommensurate magnetic order near the superconducting phase boundary in $Fe_{1+\delta}Te_{1-x}Se_x$. *Phys. Rev. B* **80**, 104506 (2009).
40. Hasan, M. Z. & Kane, C. L. Colloquium: Topological insulators. *Rev. Mod. Phys.* **82,** 3045 (2010).
41. Schubert, G. *et al*. Fate of topological-insulator surface states under strong disorder. *Phys. Rev. B* **85,** 201105 (2012).
42. Sacksteder, V., Ohtsuki, T. & Kobayashi, K. Modification and control of topological insulator surface states using surface disorder. *Phys. Rev. Applied* **3,** 064006 (2015).
43. Hao, N. & Hu, J.-P. Topological quantum states of matter in iron-based superconductors: From concepts to material realization. https://arxiv.org/abs/1811.03802 (2018).
44. Susskind, L. Lattice fermions. *Phys. Rev. D* **16,** 3031 (1977).
45. Stacey, R. Eliminating lattice fermion doubling. *Phys. Rev. D* **26,** 468 (1982).
46. Gutiérrez, C., *et al.* Interaction-driven quantum Hall wedding cake–like structures in graphene quantum dots. *Science,* **361,** 789–794 (2018).
47. Arsoski, V. V. *et al.* An efficient finite-difference scheme for computation of electron states in free-standing and core–shell quantum wires. *Comput. Phys. Commun.* **197,** 17-26 (2015).

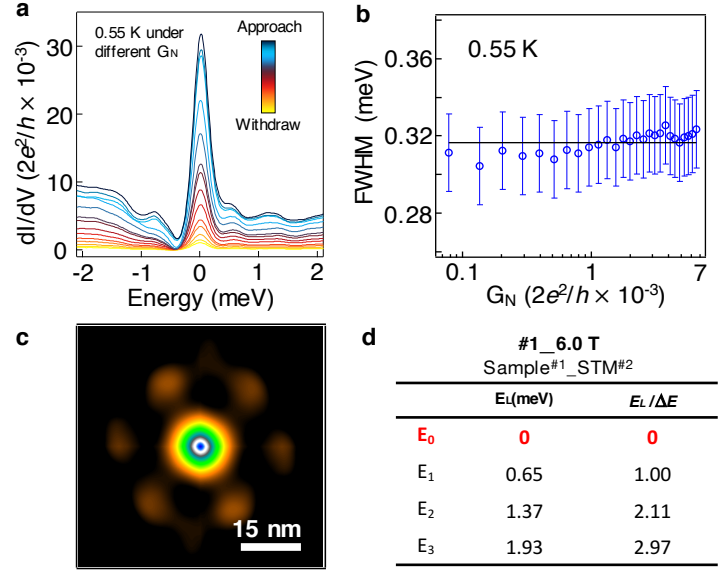

**Extended Data Fig. 1 | Characteristics of ZBCP within integer quantized CBSs. a,** Tunneling barrier evolution of the ZBCPs measured on vortex #1 shown in the main text, within integer-quantized CBSs. The ZBCPs are located at 0 meV over two orders of magnitude of tunneling barrier conductance. **b,** FWHM of ZBCPs under different tunneling barriers. The average width is about 0.32 meV. **c,** Autocorrelation of the zero bias conductance mapping shown in Fig. 1a, supporting an ordered hexagonal vortex lattice. **d,** Summary of energy positions of CBSs in vortex #1.

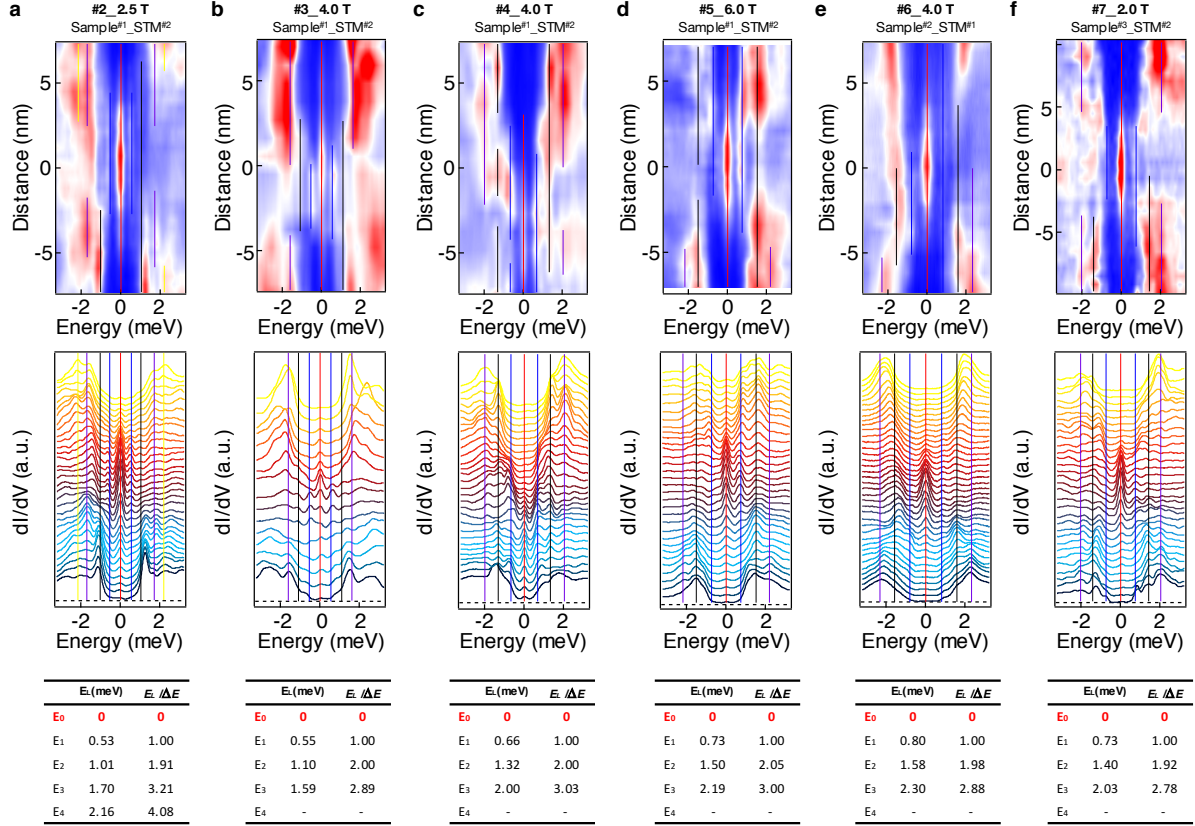

**Extended Data Fig. 2 | More examples of integer quantized CBSs. a - f,** Six topological vortices measured under different magnetic fields, samples and equipment. The first row shows intensity line-cut plots. The second row shows the corresponding waterfall-like plots. The third row lists the energy positions of CBSs. Note: **(d)** is the same data used in Fig. S3 of the previous study[17].

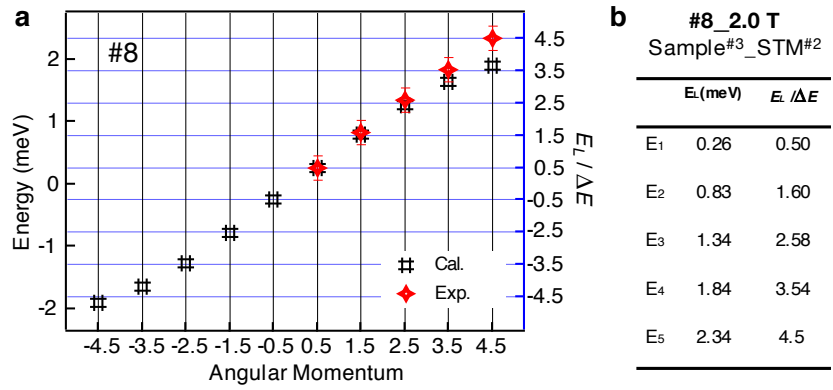

**Extended Data Fig. 3 | Numerical calculation of the ordinary vortex #8. a,** Numerical calculations of the energy eigenvalues of CBSs versus angular momentum. The red symbols are experimental data, plotting the number of energy level as the eigenvalue of angular momentum for comparison. The black symbols are numerical calculations based on solving Bogoliubov-de Gennes equation with the parabolic conventional bands. **b,** Summary of energy positions of CBSs in vortex #8.

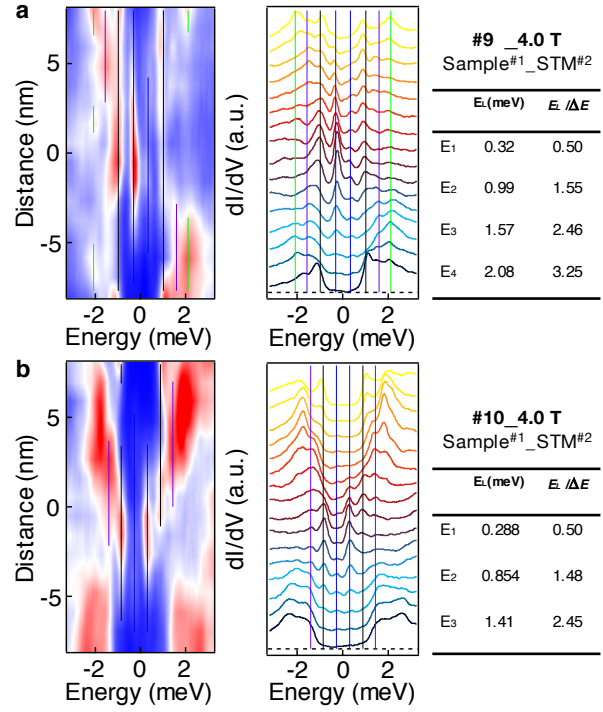

**Extended Data Fig. 4 | More examples of half-odd-integer quantized CBSs. a - b,** Vortex bound states in ordinary vortex#9 and #10, respectively. The first column shows intensity line-cut plots. The second column shows the corresponding waterfall-like plots. The third column lists the energy positions of CBSs.

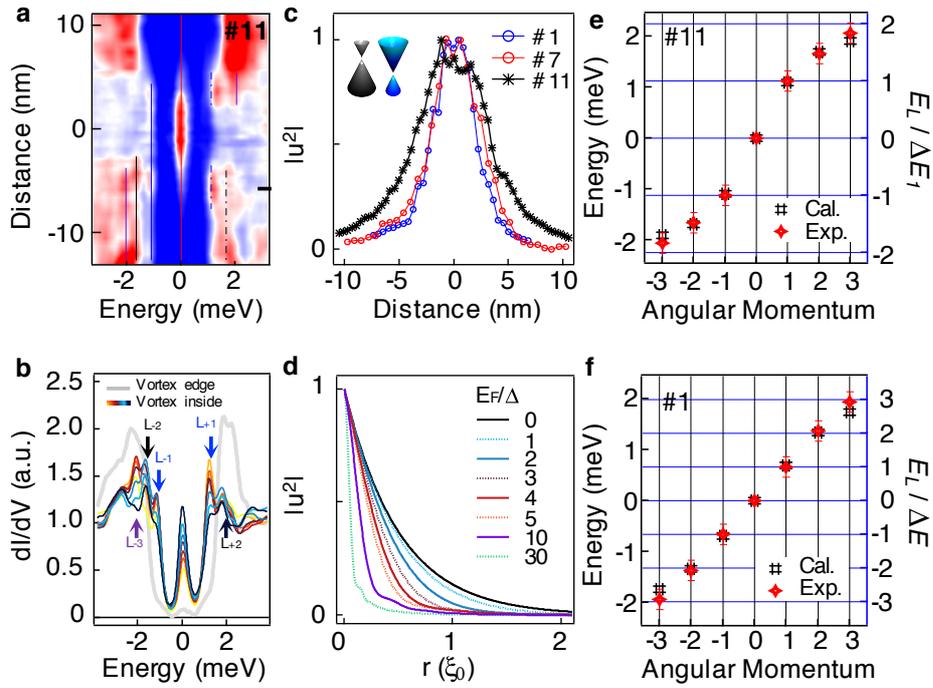

**Extended Data Fig. 5 | A topological vortex near the zero doping limit. a,** A line-cut intensity plot of the topological vortex #11, in which the discrete CBSs are neither integer quantized nor half-odd-integer quantized. **b,** Eight tunneling conductance spectra shown in color are measured around 5 nm away from vortex center as indicated by a short black bar in (**a**). A gray curve measured at the vortex edge is overlapped on (**b**) for comparison. **c,** A spatial intensity profile of the MZMs extracted from vortex#1, #7 and #11. The MZM in the integer quantized topological vortex#1 and #7 has narrower spatial distributions, while MZM in the topological vortex#11 without integer quantized property has wider spatial distributions. Insert: a cartoon demonstrating chemical potential difference of the underlying Dirac bands of vortex #1 and #11. **d,** An analytical model plot of Majorana wave function used in previous study[17]. It shows that a MZM with larger Fermi energy obtains narrower spatial distribution. **e, f,** Numerical calculations of the energy eigenvalues of CBSs versus angular momentum. The calculation in (**e**) and (**f**) are compared with the observed peak positions of vortex #11 and #1, respectively. Red symbols show the experimental data. Model parameters: $\Delta = 2.2$ meV and $E_F = 2.64$ meV for (**e**) and $\Delta = 2.2$ meV and $E_F = 3.63$ meV for (**f**). The line-cut intensity plot in (**a**) and the black curve in (**c**) are the same data used in Fig. 2 of previous study[17].

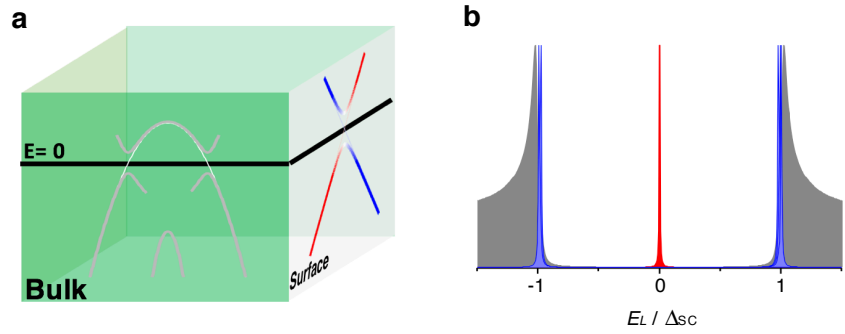

**Extended Data Fig. 6 | Topological vortices under zero doping limit. a,** Schematic plots of the surface and bulk band structure, when the underlying band structure is dominated by topological nontrivial Dirac surface state within zero doping limit. **b,** Schematic plots of the corresponding subgap CBSs. All the CBSs are pushed towards the gap edge, leaving the MZM isolated at zero energy.

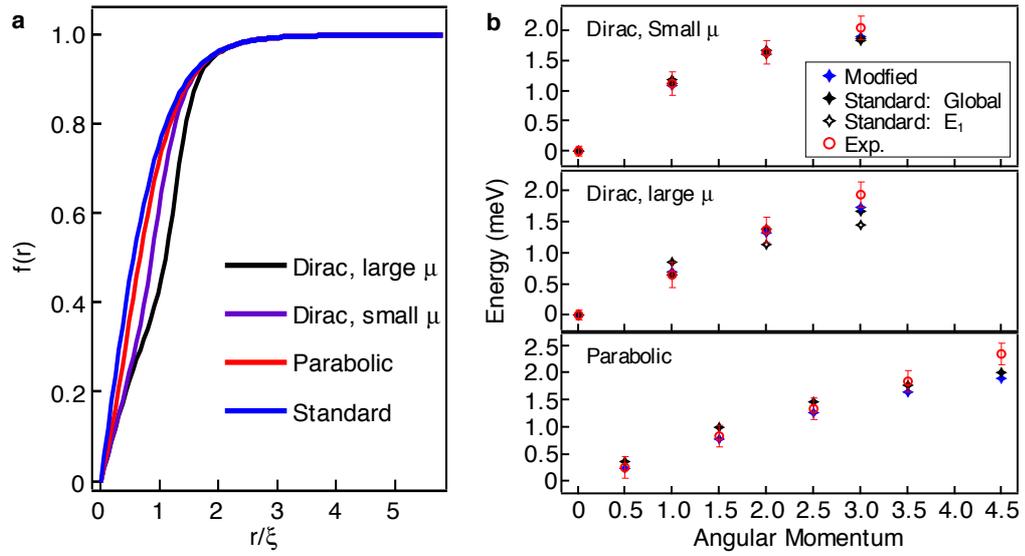

**Extended Data Fig. 7 | Details on numerical calculations. a,** Radial superconducting gap profiles of vortices used in calculations. The profile modification possibly arises from an additional dip in the superconducting order parameter due to inhomogeneities of the material. **b,** Calculation results with different methods. The blue symbols are numerical results based on modified radial profile shown in **(a)**. Red circles are experiment results. Solid black symbols are numerical results based on standard radial profile and fitting by optimizing the global deviation with experiment values. Hollow black symbols are numerical results based on standard radial profile and fitting by optimizing the $E_1$ deviation with experiment values.